\documentstyle[12pt,epsf]{article} \hoffset -0.5in \textwidth 6.00in
\textheight
8.5in  \parskip 7pt \openup1.5\jot \parindent=0.5in
\newskip\humongous \humongous=0pt plus 1000pt minus 1000pt
\def\caja{\mathsurround=0pt}
\def\eqalign#1{\,\vcenter{\openup1\jot \caja
        \ialign{\strut \hfil$\displaystyle{##}$&$
        \displaystyle{{}##}$\hfil\crcr#1\crcr}}\,}
\newif\ifdtup

\def\eqright #1\cr{\noalign{\hfill$\displaystyle{{}#1}$}}
\def\eqleft #1\cr{\noalign{\noindent$\displaystyle{{}#1}$\hfill}}

\def\oldreffmt#1{\rlap{[#1]} \hbox to 2\parindent{}}

\def\figfmt#1{\rlap{Figure {#1}} \hbox to 1in{}}

\def\sectioneq{\def\theequation{\thesection.\arabic{equation}}{\let
\holdsection=\section\def\section{\setcounter{equation}{0}\holdsection}}}%

\def\auto{\eqno(\refstepcounter{equation}\theequation)}
\def\begineq #1\endeq{$$ \refstepcounter{equation}\eqalign{#1}\eqno
	(\theequation) $$}
\def\contlimit{\,{\hbox{$\longrightarrow$}\kern-1.8em\lower1ex
\hbox{${\scriptstyle (a\rightarrow0)}$}}\,}
\def\centeron#1#2{{\setbox0=\hbox{#1}\setbox1=\hbox{#2}\ifdim
\wd1>\wd0\kern.5\wd1\kern-.5\wd0\fi
\copy0\kern-.5\wd0\kern-.5\wd1\copy1\ifdim\wd0>\wd1
\kern.5\wd0\kern-.5\wd1\fi}}
\def\centerover#1#2{\centeron{#1}{\setbox0=\hbox{#1}\setbox
1=\hbox{#2}\raise\ht0\hbox{\raise\dp1\hbox{\copy1}}}}
\def\centerunder#1#2{\centeron{#1}{\setbox0=\hbox{#1}\setbox
1=\hbox{#2}\lower\dp0\hbox{\lower\ht1\hbox{\copy1}}}}
\def\lsim{\;\centeron{\raise.35ex\hbox{$<$}}{\lower.65ex\hbox
{$\sim$}}\;}
\def\gsim{\;\centeron{\raise.35ex\hbox{$>$}}{\lower.65ex\hbox
{$\sim$}}\;}

\def\super#1{\ifmmode \hbox{\textsuper{#1}}\else\textsuper{#1}\fi}
\def\textsuper#1{\newcount\holdspacefactor\holdspacefactor=\spacefactor
$^{#1}$\spacefactor=\holdspacefactor}

\def\getcite#1,{\advance\citenumber by1
\def\getcitearg{#1}\def\lastarg{@}
\ifnum\citenumber=1
\ref{#1}\let\next=\getcite\else\ifx\getcitearg\lastarg\let\next=\relax
\else ,\ref{#1}\let\next=\getcite\fi\fi\next}

\def\pom{{\rm P\kern -0.53em\llap I\,}}
\def\spom{{\rm P\kern -0.36em\llap \small I\,}}
\def\sspom{{\rm P\kern -0.33em\llap \footnotesize I\,}}

\relax

\newskip\humongous \humongous=0pt plus 1000pt minus 1000pt
\def\caja{\mathsurround=0pt}
\def\eqalign#1{\,\vcenter{\openup1\jot \caja
        \ialign{\strut \hfil$\displaystyle{##}$&$
        \displaystyle{{}##}$\hfil\crcr#1\crcr}}\,}
\newif\ifdtup

\def\eqright #1\cr{\noalign{\hfill$\displaystyle{{}#1}$}}
\def\eqleft #1\cr{\noalign{\noindent$\displaystyle{{}#1}$\hfill}}

\def\oldreffmt#1{\rlap{[#1]} \hbox to 2\parindent{}}

\def\figfmt#1{\rlap{Figure {#1}} \hbox to 1in{}}

\def\auto{\eqno(\refstepcounter{equation}\theequation)}
\def\begineq #1\endeq{$$ \refstepcounter{equation}\eqalign{#1}\eqno
	(\theequation) $$}
\def\contlimit{\,{\hbox{$\longrightarrow$}\kern-1.8em\lower1ex
\hbox{${\scriptstyle (a\rightarrow0)}$}}\,}
\def\centeron#1#2{{\setbox0=\hbox{#1}\setbox1=\hbox{#2}\ifdim
\wd1>\wd0\kern.5\wd1\kern-.5\wd0\fi
\copy0\kern-.5\wd0\kern-.5\wd1\copy1\ifdim\wd0>\wd1
\kern.5\wd0\kern-.5\wd1\fi}}
\def\centerover#1#2{\centeron{#1}{\setbox0=\hbox{#1}\setbox
1=\hbox{#2}\raise\ht0\hbox{\raise\dp1\hbox{\copy1}}}}
\def\centerunder#1#2{\centeron{#1}{\setbox0=\hbox{#1}\setbox
1=\hbox{#2}\lower\dp0\hbox{\lower\ht1\hbox{\copy1}}}}
\def\lsim{\;\centeron{\raise.35ex\hbox{$<$}}{\lower.65ex\hbox
{$\sim$}}\;}
\def\gsim{\;\centeron{\raise.35ex\hbox{$>$}}{\lower.65ex\hbox
{$\sim$}}\;}

\def\super#1{\ifmmode \hbox{\textsuper{#1}}\else\textsuper{#1}\fi}
\def\textsuper#1{\newcount\holdspacefactor\holdspacefactor=\spacefactor
$^{#1}$\spacefactor=\holdspacefactor}

\def\getcite#1,{\advance\citenumber by1
\ifnum\citenumber=1
\ref{#1}\let\next=\getcite\else\ifx#1@\let\next=\relax
\else ,\ref{#1}\let\next=\getcite\fi\fi\next}

\def\upon #1/#2 {{\textstyle{#1\over #2}}}
\relax

\sectioneq

\def\til#1{\centeron{\hbox{$#1$}}{\lower 2ex\hbox{$\char'176$}}}
\def\tild#1{\centeron{\hbox{$\,#1$}}{\lower 2.5ex\hbox{$\char'176$}}}
\def\sumtil{\centeron{\hbox{$\displaystyle\sum$}}{\lower
-1.5ex\hbox{$\widetilde{\phantom{xx}}$}}}

\def\pom{{\rm P\kern -0.53em\llap I\,}}
\def\spom{{\rm P\kern -0.36em\llap \small I\,}}
\def\sspom{{\rm P\kern -0.33em\llap \footnotesize I\,}}

\newcommand{\bit}{\begin{itemize}}
\newcommand{\eit}{\end{itemize}}

\newcommand{\beq}{\begin{equation}}
\newcommand{\eeq}{\end{equation}}
\newcommand{\beqa}{\begin{eqnarray}}
\newcommand{\eeqa}{\end{eqnarray}}

\begin{document}

\begin{titlepage}

\rightline{\vbox{\halign{&#\hfil\cr
&ANL-HEP-CP-95-78\cr
&UF-IFT-HEP-95-19\cr}}}

\vspace{.4in}

\begin{center}

{\large\bf
HIGHER-ORDER CORRECTIONS TO BFKL EVOLUTION FROM $t$-CHANNEL UNITARITY}
\footnote{Presented by Alan R. White at VIth International Conference on
Elastic and Diffractive Scattering ``Frontiers in Strong Interactions'',
Blois, France (June 1995). }
\footnote{Work supported by the U.S. Department
of Energy, Division of High Energy Physics, Contracts
W-31-109-ENG-38 and DEFG05-86-ER-40272}

\medskip

{Claudio Corian\`{o}$^{a,b}$\footnote{
coriano@phys.ufl.edu ~$^{\#}$arw@hep.anl.gov}
and \ Alan. R. White$^{a\#}$}

\vskip 0.6cm

\centerline{$^a$High Energy Physics Division}
\centerline{Argonne National Laboratory}
\centerline{9700 South Cass, Il 60439, USA.}
\vspace{0.5cm}

\centerline{$^b$Institute for Fundamental Theory}
\centerline{Department of Physics}
\centerline{ University of Florida at Gainesville, FL 32611, USA}
\vspace{0.5cm}

\end{center}

\begin{abstract}
Using reggeon diagrams as a partial implementation of $t$-channel unitarity,
$O(g^4)$ corrections to the BFKL evolution equation have been obtained. We
describe the spectrum and holomorphic factorization properties of the resulting
scale-invariant kernel. For a gauge theory, $t$-channel unitarity can be
studied directly in the complex $j$-plane by implementing Ward identity
constraints together with the group structure of reggeon interactions. We
discuss how both the $O(g^2)$ BFKL kernel and the $O(g^4)$ corrections can
then be derived.

\end{abstract}

\end{titlepage}

Currently the most familiar application of the BFKL equation is to the
evolution of parton distributions at small-x. The solution
$F(x,k^2) \sim x^{1- \alpha_0}$, with $\alpha_0 \sim 1.5$, is
well-known. In the non-forward direction the equation becomes a
``reggeon Bethe-Salpeter equation" i.e.
$$
\eqalign{ F(\omega,k,q-k) ~=~\tilde{F}+
{1 \over (2\pi)^3}\int {d^2k' \over
(k')^2(k'-q)^2}~&\Gamma_2(\omega,k',q-k')\cr
\times &\tilde{K}(k,k',q) F(\omega,k',q-k') }
\auto
$$
$\omega~(=j-1)$ is conjugate to $ln[{1 \over x}]$, $K$ is a 2-2 reggeon
interaction and $\Gamma_2$ is a two-reggeon propagator. To obtain higher-order
corrections we must study higher-order reggeon interactions. We can
do this directly via high-energy ``$s$-channel'' unitarity calculations.
{\it Alternatively} we can sew reggeon amplitudes together in the
$j$-plane via ``$t$-channel'' unitarity.

Initially$^{\cite{ker}}$ we suggested sewing reggeon amplitudes together
by using reggeon diagrams. With a ``nonsense zero'' in the
three-reggeon vertex, many reggeon singularities in the diagrams are cancelled,
leaving only particle singularities generating leading and next-to-leading
order reggeon interactions. We obtained a
leading-order 2-4 kernel and the $O(g^4)$ higher-order 2-2 kernel
$K^{4n}$, written in terms of transverse momentum diagrams in Fig.~1

\begin{center}

\leavevmode
\epsfxsize=5in
\epsffile{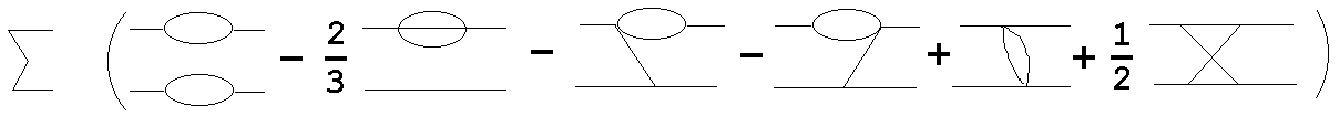}

Fig.~1 The diagrammatic representation of $K^{4n}$.
\end{center}
$K^{4n}$ is infra-red finite and satisfies the Ward identity
constraints of gauge invariance.

The last diagram of Fig.~1 is the
most difficult to evaluate since it contains the box diagram
$$
\eqalign{I_4 (p_1&,p_2,p_3,p_4,m^2)~=~\int d^2p ~\Pi_{i=1}^4 ~
{1\over [(p-p_i)^2 - m^2] }}
$$
We evaluated$^{\cite{cw}}$ $I_4$ as a sum of logarithms, i.e.
$$
I_4=\sum_{j\,\,<\,\,k}A_{jk} F_{jk}~
\auto
$$
where the $A_{jk}$ are ``tree-diagrams'' obtained by putting internal lines
$j$ and $k$ on-shell and
$$
F_{jk}~=~{i\pi\over \lambda^{1/2}(p_{jk}^2,m^2,m^2)}Log\Biggl[
{p_{jk}^2 -2 m^2 -\lambda^{1/2}(p_{jk}^2,m^2,m^2)\over
p_{jk}^2 -2 m^2 + \lambda^{1/2}(p_{jk}^2,m^2,m^2)}\Biggr]
\auto
$$
with
$$
p_{jk}=(p_j-p_k)^2.
\auto
$$

In the forward direction the $A_{jk}$ simplify considerably and
if $K_{BFKL}$ is the leading-order BFKL kernel we find$^{\cite{cw}}$ we can
write
$$
K^{(4n)}(k,k')~=~{1 \over 4} (K_{BFKL})^2~+~{\cal K}_2
\auto
$$
where
$$
{\cal K}_2~~=~{1 \over
2\pi^2 } ~ {k^2 {k'}^2 (k^2-{k'}^2)\over
(k+k')^2 (k-k')^2}~Log\Biggl[{k^2 \over {k'}^2} \Biggr]
\auto
$$
${\cal K}_2$ has a number of attractive properties. It is separately
infra-red finite and has a spectrum very reminiscent of the leading-order
kernel. We anticipate that eventually it will be established as the forward
component of a well-defined {\it conformally invariant} $O(g^4)$
contribution to the BFKL kernel.

Using the orthogonal eigenfunctions
$$
\phi_{\nu,n}(k)~=~({k}^2)^{1/2 + i\nu}~e^{i n \theta}
\auto
$$
the eigenvalues of ${\cal K}_2$ are
$$
\eqalign{ \Lambda(\nu,n) ~=~-~{1 + (-1)^n\over 8\pi}
\biggl(\beta'\bigl({|n| + 1\over 2} +
i\nu\bigr)
{}~+~\beta'\bigl({|n| + 1 \over 2} -i\nu\bigr)\biggr). }
\auto
$$
where
$$
\beta'(x)~=~{1\over 4}\biggl(\psi'\bigl({x+1\over 2}\bigr) -
\psi'\bigl({x\over 2}\bigr)\biggr)
\auto
$$
with
$$
\psi'(z)~=~\sum_{r=0}^{\infty} {1
\over (r+z)^2}.
\auto
$$
Note that
$$
\eqalign{ \Lambda(-\nu,-n) - \Lambda(\nu,n) &\sim
\sum_{t=-n/4}^{n/4~-1}\Biggl[ {1 \over (t + {3 \over 4} - {i\nu \over 2})^2}-
{1 \over (t  + {1 \over 4} - {i\nu \over 2})^2}\cr
&~~~~~~~~~+ {1 \over (t + {3 \over 4} + {i\nu \over 2})^2}
- {1 \over (t+ {1\over 4} + {i\nu \over 2})^2} \Biggr]\cr
&=~~0}
\auto
$$
{\it for $n$ even}. As a result we can write
$$
\Lambda(\nu,n)~=~{\cal G}\bigl[m(1-m)\bigr]~+
{}~{\cal G}\bigl[\tilde{m}(1-\tilde{m})\bigr]
\auto
$$
where
$$
m=1/2 + i\nu + n/2~~~ and ~~~\tilde{m}= 1/2 + i\nu -n/2
\auto
$$
are conformal weights. This is the property of
{\it holomorphic factorization} which generally accompanies conformal
invariance.

Moving on to numerical results, we note that $\Lambda(0,0)$ is
the leading eigenvalue. With a simple reggeon diagram
normalization$^{\cite{cw}}$, the correction to $\alpha_0$ is given by
$$
\eqalign{ {9g^4\over 16\pi^3}\Lambda(0,0)
{}~&=~-~{9g^4 \over 32\pi^4 }\beta'(1/2)\cr
{}~&\sim~-16.3 {{\alpha_s}^2 \over \pi^2}\cr
{}~&\sim~ - 0.06}
\auto
$$
However, $K^{4n}(q,k,k')$  contains disconnected diagrams (the first term
in Fig.~1) which can not be interpreted in terms of reggeization
effects. Eliminating these diagrams, while retaining scale-invariance,
leads uniquely to
$$
\tilde{K}^{(4)} =K^{(4n)}-[{K}_{BFKL}]^2
\auto
$$
as a consistent scale-invariant $O(g^4)$ kernel. In this case the
modification of $\alpha_0$ is ($\chi(0,0)$ is the $O(g^2)$
eigenvalue)
$$
\eqalign{{9g^4 \over 16 {\pi}^4}&\biggl(
-3 [\chi(0,0)]^2
{}~-~\Lambda(0,0) \biggr)\cr
{}~&\sim ~-68 {{\alpha_s}^2 \over \pi^2}\cr
{}~&\sim~-0.25}
\auto
$$
indicating that higher-order corrections may give a substantial negative
correction to the leading-order BFKL result.

Unfortunately, several questions related to the significance of the numerical
results are left unanswered by the reggeon diagram construction. In particular,
\begin{itemize}
\item{what is the justification for the reduction to transverse momentum
diagrams? }
\item{How do scales enter and what is the significance of conformal
invariance?}
\end{itemize}
A related but more fundamental approach to the derivation of
reggeon interactions, which potentially can answer such questions, is
provided by the analytic continuation of multiparticle unitarity equations
in the $j$-plane. This is a powerful formalism extensively based on
multiparticle dispersion theory. A full description of the application to
gauge theories can be found in$^{\cite{cw2}}$. The essential elements are
\begin{itemize}
\item {Gauge invariance is input via the Ward identity constraint that
reggeon interaction vertices vanish when any reggeon transverse momentum
goes to zero.}
\item {The ``nonsense'' zero/pole structure required by general
analyticity properties is imposed. }
\item {The group structure is input via the triple reggeon vertex.}
\item {$t$-channel unitarity is used to determine both $j$-plane
Regge cut discontinuities and particle threshold discontinuities due to
``nonsense'' states.}
\item {The $j$-plane and $t$-plane discontinuities are expanded
around $j=1$ and in powers of $g^2$.}
\end{itemize}

The most important element is the {\it nonsense state particle
discontinuities.} Unitarity dictates that $j$-plane reggeon discontinuities
are given by transverse momentum integrals. For particle discontinuities
this is the case only when special kinematic circumstances determine that
only nonsenses states are involved. When this happens, expansion around $j=1$
(the equivalent of expanding in powers of logs in momentum space), and in
powers of $g^2$, straightforwardly gives interactions etc. in terms of
transverse momentum diagrams. The $O(g^2)$ contribution to the trajectory
function arises from the two-particle discontinuity of the reggeon
propagator. The 2-2 reggeon interaction, i.e. the BFKL kernel, is extracted
from the nonsense-state discontinuities of the two reggeon propagator Green
function illustrated in Fig.~2.
\begin{center}
\leavevmode
\epsfxsize=5in
\epsffile{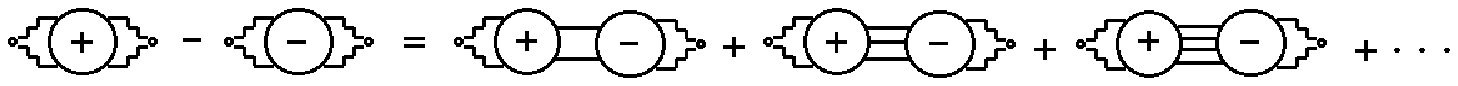}

Fig.~2. Nonsense state discontinuities of the two reggeon propagator Green
function
\end{center}
The $O(g^2)$ kernel is unambiguously derived from the
three-particle discontinuity as $q^2 \to 0$ since {\it only nonsense states are
involved}. Contributions to the $O(g^4)$ kernel that we have discussed above
are from the four-particle state. More elaborate kinematic constraints are
necessary for the reduction to transverse momentum diagrams and a number of
qualifications of the reggeon diagram results emerge.

The most attractive part of the $O(g^4)$ kernel, ${\cal K}_2$, can be
derived directly when {\it both} $q^2 \to 0$ and either $k^2 \to 0$, or
${k'}^2 \to 0$, {\it apart from an
overall normalization factor}. The
existence of all but the first term in Fig.~1 follows once the existence of
a 1-3 reggeon vertex is assumed. (We have noted above that the first term in
Fig.~1 is removed by the introduction of the square of the leading-order
kernel). The reduction to transverse
momentum integrals is again only valid for $q^2,k^2~\to 0$, or
$q^2,{k'}^2~\to 0$. These results are
consistent with those obtained by Kirschner$^{\cite{kir}}$ from the
$s$-channel multi-Regge effective lagrangian formalism. It seems possible
that the scale-dependence can be built up by adding internal logarithms to
the transverse momentum integrals while maintaining the Ward identity
constraints.

Finally we note that both $t$-channel unitarity and the multi-Regge
effective lagrangian imply that the introduction of scales will modify the
normalization and significantly modify the kernel at large $q^2,k^2,{k'}^2$.
When the full $O(\alpha_s^2)$ kernel is calculated we hope the comparison
will show how the reggeon diagram formalism can usefully approximate yet
higher-order contributions.


\begin{thebibliography}{99}

\bibitem{ker} A.~R.~White, {\it Phys. Lett.} {\bf B334}, 87 (1994).

\bibitem{cw}  C.~Corian\`{o} and A.~R.~White {\it Phys. Rev. Lett.}{\bf 74},
4980 (1995), {\it Nucl. Phys.} {\bf B451}, 231 (1995).

\bibitem{cw2} C.~Corian\`{o} and A.~R.~White, ANL-HEP-PR-95-19.

\bibitem{kir} R.~Kirschner, LEIPZIG-18-1995, hep-ph/9505421.

\end{thebibliography}
\end{document}